# Improving Congestion Control for Concurrent Multipath Transfer through Bandwidth Estimation based Resource Pooling

Samar Shailendra, R. Bhattacharjee, *Member IEEE*, Sanjay K. Bose, *Senior Member IEEE*
Department of Electronics & Electrical Engineering, I.I.T. Guwahati, Guwahati, INDIA
*{samar, ratnajit, skbose}@iitg.ernet.in*

*Abstract*—Stream Control Transmission Protocol (SCTP) was introduced in 2001 as a multipath variant to traditional transport protocols, i.e. Transmission Control Protocol (TCP) and User Datagram Protocol (UDP). Concurrent Multipath Transfer (CMT) has been proposed as an extension for SCTP to support concurrent usage of available multiple paths. In this paper, we propose a new congestion control algorithm for CMT-SCTP based on the principle of resource pooling. We use the connection bandwidth estimates to obtain the collection of the network resources being used by different flows on multiple paths. Based on these bandwidth estimates, we have used the bandwidth estimation based resource pooling approach to adjust the congestion window of the respective paths. We compare our proposed scheme with CMT-SCTP through *ns*-2 based simulations.

I. INTRODUCTION

Considerable research work is currently underway to support multipath transport [1-7] in networks so that resources can be efficiently utilized for data transfers. In 2001, IETF proposed the Stream Control Transmission Protocol (SCTP) to support multi-homed end nodes [8-10] in a network. SCTP also has the potential to support multi-streaming though that is not part of the existing standards. This would reduce the Head of Line (HOL) blocking that is sometimes experienced by TCP [11]. Janardhan et al. [12] proposed a Concurrent Multipath Transfer (CMT) extension to SCTP which allows concurrent usage of multiple disjoint end-to-end paths in the network.

Implementation of proper Congestion Control algorithms is necessary for the operation of any transport protocol over a network and must be incorporated in any Internet based implementation. The congestion control mechanism proposed for CMT is largely based on the mechanism followed by traditional TCP and is not effective when multiple paths are concurrently used. While proposing and implementing new congestion control mechanisms for SCTP, a primary requirement would be that the new protocol implementation should not take undue advantage of the network and should not harm other fellow TCP and UDP flows with which it coexists in the network.

For providing better QoS to an end user, load balancing has become a de-facto implementation strategy for any service provider. The basic requirement for load balancing is to aggregate all the available resources. Thus load balancing is an example of resource pooling. The basic philosophy of resource pooling is to look at all the available resources as a single pool of resources. The available bandwidth of the individual paths is a typical example of the resource under consideration. According to resource pooling principle, the total bandwidth of the set of available paths will be shared equally by the flows using those paths. Currently IETF is working on MultiPath TCP, a TCP extension to support multiple path usage in parallel fashion. Raiciu et al. [13] have proposed a congestion control mechanism based upon resource pooling principle [14]. This mechanism tries to achieve the following three goals: *i*) *Improved throughput* i.e. multipath flow should perform at least as good as the best of the path available in the pool; *ii*) *Do not harm* i.e. other fellow flows on the shared link should not unduly suffer; *iii*) *Balance Congestion* i.e. the congestion experienced by all the paths should be similar.

Besides improving the congestion control algorithm, other efforts are also underway to improve the performance of multipath flows. Dreibholz et al. [15] have proposed scheduling of packets on multiple paths in an optimized fashion to improve the throughput of the network and also provide for RTT compensation on multiple paths with dissimilar RTTs. Tsai et al. [16] have proposed a multipath transmission control scheme (MTCS) to improve throughput for real time data transmission using multiple paths. They have proposed a packet scheduling policy to handle out-of-order packets in multipath transmission. Key et al. [17] have given a mathematical analysis of the benefits of doing coordinated congestion control with active state information of the paths. In this paper, we have proposed a modified congestion control algorithm for CMT-SCTP. Our proposed algorithm applies the resource pooling principle based upon the bandwidth estimates obtained by observing the data flow on the paths.

The rest of the paper is organized as follows. In Section II, we present the changes being proposed in the congestion control algorithm. Section III presents the network model of the scenarios where the proposed changes are tested and compares the results obtained with those obtained by using the CMT protocol. Section IV concludes the paper followed by references used.

## II. PROPOSED CHANGES IN CONGESTION CONTROL

While technologies like ATM can ensure guaranteed QoS to the end users, Internet typically supports a best effort delivery service. TCP and UDP are the two transport protocols that are mainly used in the current Internet. UDP is a datagram based protocol which does not provide any reliability. TCP does end-to-end error correction and provides reliability to the underlying Internet Protocol (IP) layer which is connectionless and unreliable in nature. Recently, CMT-SCTP and Multipath TCP (MPTCP) have been proposed to use multiple parallel paths for data transfer. This should provide better resource utilization and increase the error resilience of the underlying network. The congestion control of SCTP is based on the TCP congestion control. TCP based congestion control works well with standard SCTP because in spite of being a multihoming protocol, it uses only one path at a time and switches to an alternate path only if the primary path fails. However, CMT-SCTP uses all the paths concurrently. Hence it takes undue advantage of shared links on multiple paths which will create issue of fairness and TCP friendliness. In literature, it has been demonstrated that CMT flows unduly harm other network flows on the shared links and captures more bandwidth in the ratio of the number of concurrent paths being established across any shared bottleneck link en-route. These problems motivated us to reexamine the congestion control algorithm of CMT-SCTP and suggest an approach which will allow multiple paths to be used without incurring these difficulties. In this paper, we have proposed a modified congestion control algorithm using **B**andwidth **E**stimation based **R**esource **P**ooling (**BERP**) to overcome the above said shortcomings in CMT congestion control.

Wischik et al. introduced the concept of resource pooling in [14]. This principle considers all the independent resources available on the network as a single pool of resources. Recently, Raiciu et al. [13] have investigated fully coupled and link coupled congestion control algorithms for Multipath TCP (MPTCP). In this work they demonstrate that linked congestion control based algorithm exhibits better performance in terms of throughput and fairness on paths with dissimilar round trip time (RTT). However, they still use a TCP like approach to decrease the congestion window. This difference in policy for congestion window evolution causes unfairness. Hence to balance the increase in congestion window with the decrease of the same at equilibrium, a factor of aggressiveness ($\alpha$) was introduced. Dreibholz et al. [18] have compared an MPTCP-like algorithm with CMT-SCTP. In this paper, we also consider the resources being used by CMT as a single pool of resources. However, we argue that this approach is not optimum on heterogeneous networks with wireless links. Unlike wired links, losses in wireless links may happen because of reasons other than congestion (e.g. noise or link quality). Hence we consider the resources to be as a single pool of resources during the congestion detection phase as well.

Currently the CMT congestion control algorithm is based upon the standard TCP based congestion control algorithm with some minor modifications. Note that TCP's congestion control algorithm consists of three phases, i) Slow Start phase, ii) Congestion Avoidance phase and iii) Congestion detection phase. CMT starts the connection with the slow-start phase. It exhibits exponential growth in the congestion window till the slow start threshold (*ssthresh*) is reached. Once the *cwnd* value crosses *ssthresh*, it gets into the congestion avoidance phase. During this congestion avoidance phase, CMT increases its congestion window by one Maximum Transmission Unit (MTU) for every RTT. We propose to modify congestion avoidance rule as follows.

*Congestion Avoidance Phase:*

Increase the value of cwnd of $i^{th}$ path by:

- min($\alpha * P_a * MTU/w_T$, $P_a * MTU/w_i$)
- $P_a = P_a - w_i$

where

$P_a$ = Partial Bytes Acked

$w_T$ = Total congestion window

$w_i$ = Congestion window of $i^{th}$ path

$srtt_i$ = smoothed RTT of $i^{th}$ path

$$\alpha = \frac{2 * w_T * \max\left(\dfrac{\beta_i * w_i}{srtt_i^2}\right)}{\left(\sum \dfrac{w_i}{srtt_i}\right)^2} \quad (1)$$

We have adopted the same methodology to calculate $\alpha$, the factor of aggressiveness as the one suggested in [13]. The value of $\beta$ is given by (2).

Whenever the sender experiences any loss in packet, traditional SCTP algorithm infers that the path is in congestion and to reduce the congestion on that path it cuts the congestion window by half. However, in wireless links this may not always be the case. A wireless link may experience higher packet loss even when the link is not in congestion. It has also been demonstrated in literature that CMT may be unfair to fellow TCP flows on a shared link and may capture unduly more bandwidth. We plan to avoid this in our approach by using resource pooling during congestion detection for adjusting the congestion window. To apply resource pooling, we need to measure the available resources, i.e. bandwidth used by the flow. We use the bandwidth estimation for measuring the link bandwidth available to the user. Several techniques for bandwidth estimation have been proposed in literature. We use the same approach for the bandwidth estimation as the one being used in TCP Westwood.

The modified rule for decreasing the congestion window will be as follows.

*Congestion Detection Phase:*

if (four duplicates are received)

- $s_i = max(w_i – \beta_i * w_i , 4 * MTU)$
- $w_i = s_i$

if (Timeout occurred)

- $s_i = max(w_i – \beta_i * w_i , 4 * MTU)$

- $w_i = 1 * MTU$

where

$s_i$ = slow start threshold of $i^{th}$ path

$$\beta_i = \frac{BWE_i}{\sum_i BWE_i} \quad (2)$$

where, $BWE_i$ = Bandwidth estimate of $i^{th}$ path

Since CMT uses all the paths concurrently, we can use the SACKs on each path for measuring the bandwidth estimate on respective path. If the $k^{th}$ SACK is received at time instant $t_k$ and acknowledges $d_k$ new bytes then the current bandwidth sample $B_k$ is given by

$$B_k = \frac{d_k}{(t_k - t_{k-1})} \quad (3)$$

In order to average the sampled bandwidth, a low pass filter is applied to these samples as in [19]. The filtered bandwidth estimate is given as

$$\hat{B}_k = p.\hat{B}_{k-1} + (1-p).\left(\frac{B_k + B_{k-1}}{2}\right) \quad (4)$$

where $p$ is a constant whose value is typically set to 0.9. Note that MPTCP congestion control is also a special case of our algorithm with $\beta$=0.5.

III. PERFORMANCE ANALYSIS

To study the performance of the proposed congestion control, we have considered three scenarios.

A. *Disjoint Paths with no congestion*

We consider the network graph as shown in Fig.1. This network consists of a dual homed source (S) and a dual-homed destination (D). Both the paths have several disjoints links as shown in the figure. The links between the source interfaces and the Node-1 (N1) and Node-2 (N2) are wired links with bandwidth 10Mpbs and fixed packet loss probability of 1%. The links between N1 & Node-3 (N3) and also N2 & Node-4 (N4) are links with 1Mbps bandwidth each and the packet loss probability of these links is varied in the range 1-10%. The links between destination node interface and N3 and N4 are also wired links with 10Mbps bandwidth and 1% packet loss probability.

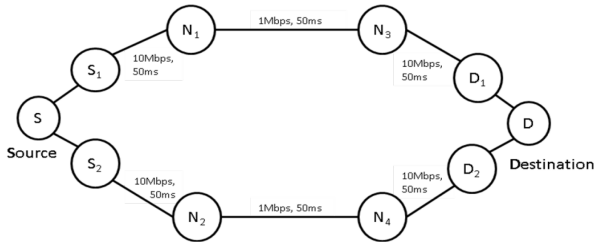

Figure 1. Network Graph for disjoint path with similar characteristics

For our tests, we have assumed that the source is transferring a 60MB file to the destination using *ftp*. We have measured the time taken to transfer the file from the source to the destination while the packet loss probability of links is varied from 1% to 10% (Fig.2) for both the CMT congestion control (CMT-CC) algorithm and our proposed algorithm (CMT-BERP). The results indicate that when the losses are purely due to error and there is no congestion in the links, CMT-BERP algorithm performs better in comparison to the CMT-CC algorithm. This is because CMT-BERP algorithm does not abruptly decrease the congestion window on every loss by assuming that this is being caused by congestion. Instead, it estimates the bandwidth of the connection and decreases the bandwidth according to the resource pooling principle. In Fig. 3, we show how the congestion window at the source evolves for the case when the packet loss probability is 10%.

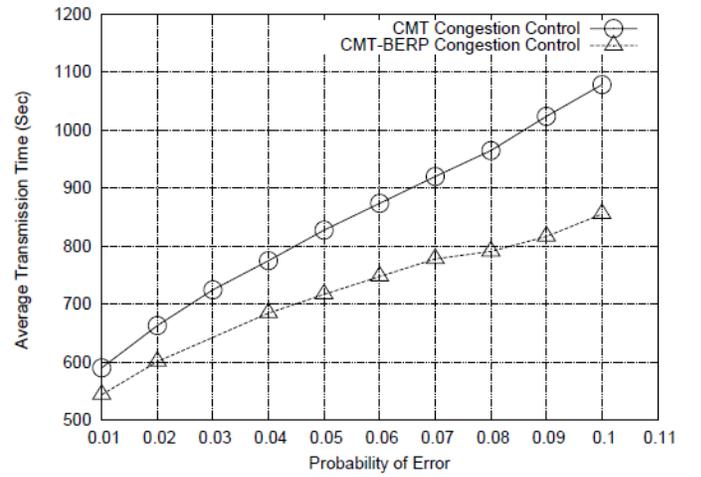

Figure 2. Average Transmission Time for disjoint paths

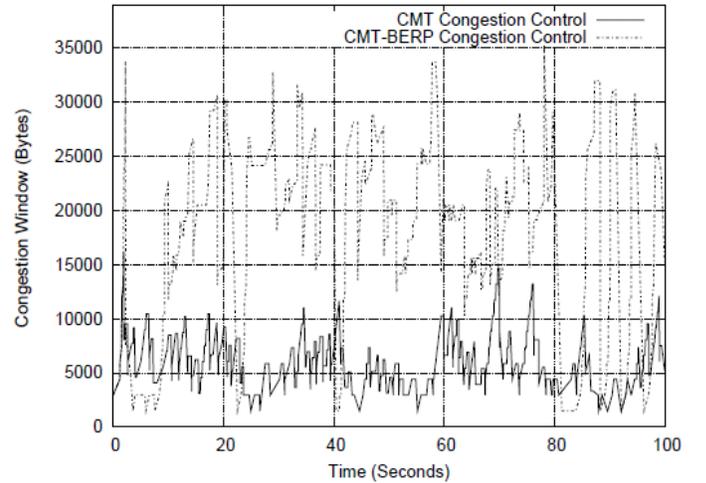

Figure 3. Congestion Window evolution for CMT congestion control and CMT-BERP algorithm for packet loss probability of 10%.

## B. Paths with Shared Bottleneck link

Fig. 4 shows an example of a scenario where there is a shared bottleneck link. The link characteristics are same as in (A) except that the shared bottleneck link between N1 and N2 has 1Mpbs bandwidth with a packet loss probability that is varied from 1% to 10%.

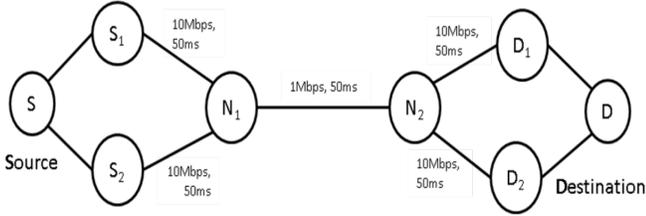

Figure 4. Network Graph for shared bottleneck

The test results for transferring a 60MB file are shown in Fig.5 which shows the average file transmission time for varying packet loss probabilities in the bottleneck link.

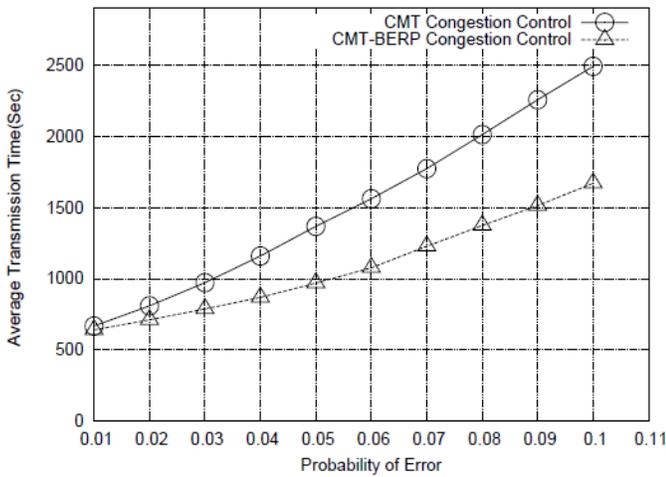

Figure 5. Average Transmission Time for Shared bottleneck link.

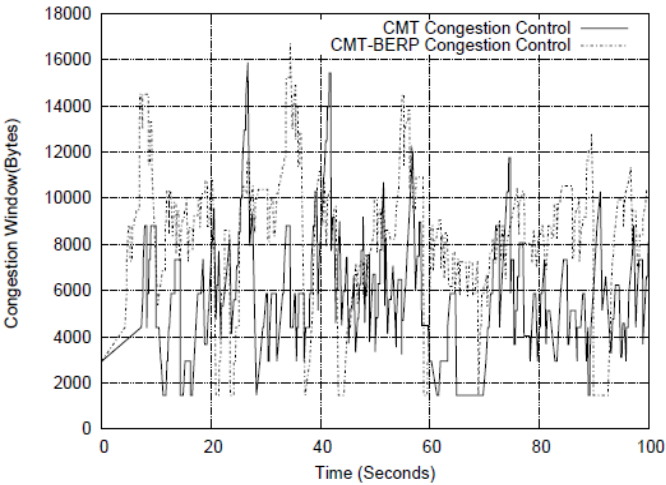

Figure 6. Congestion Window evolution for CMT congesion control and CMT-BERP algorithm for packet loss probability of 10%.

The results indicate that CMT-BERP algorithm performs better than the CMT-CC algorithm for this scenario. However, because of the shared bottleneck link, the average transmission time is higher than the earlier case where disjoint paths were used. The way the Congestion Window evolves in this scenario has been shown in Fig. 6 for both the congestion control schemes when the bottleneck link has 10% packet loss probability.

## C. Disjoint paths with only congestion losses

For this scenario, as in Fig. 7, we consider links between N1, N2 and N3, N4 to be error free. We assume that U1 and U2 are UDP sources generating CBR traffic with U3 and U4 as their respective sink nodes. These UDP sources are introduced to create congestion in the network between the source S and destination D.

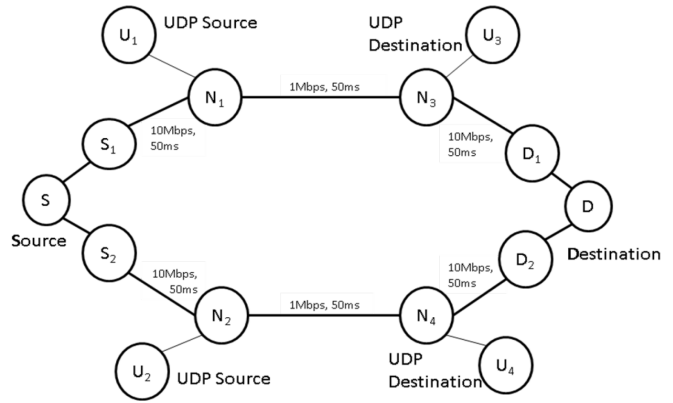

Figure 7. Network Graph for disjoint links with congestion losses

The packet size assumed for these UDP sources is 512 bytes. In this case, we have obtained average throughput achieved for both CMT-CC algorithm and CMT-BERP algorithm. The results are shown in Fig. 8.

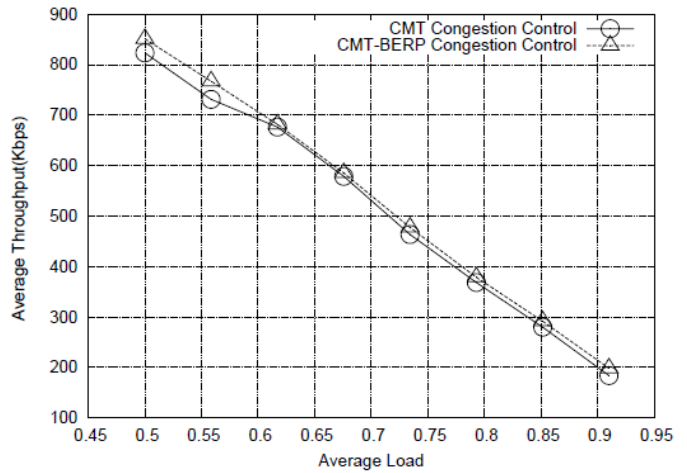

Figure 8. Average Throughput for different Average load of UDP traffic.

We observe that in this scenario, both congestion control strategies show almost identical performances. This is expected, as in this system, virtually all the losses are caused by congestion. The evolution of the congestion window for this scenario has also been shown in Fig. 9 for average UDP traffic load of 0.9.

IV. CONCLUSION

We have proposed and studied the performance of a modified congestion control algorithm. This algorithm is based on the multipath TCP congestion control algorithm. Though we have tested it only for CMT-SCTP, we expect that this can be suitably adapted for use with any multipath transport protocol. Our proposed algorithm (CMT-BERP) uses resource pooling during congestion detection phase for better overall performance where the bandwidth resource available is obtained using a suitable bandwidth estimation approach. The results indicate that CMT-BERP algorithm adapts better to packet drops than CMT congestion control algorithm and therefore provides better resource utilization. Since this algorithm is based upon the MPTCP algorithm, it is likely to inherit the fairness and friendliness property of MPTCP as well.

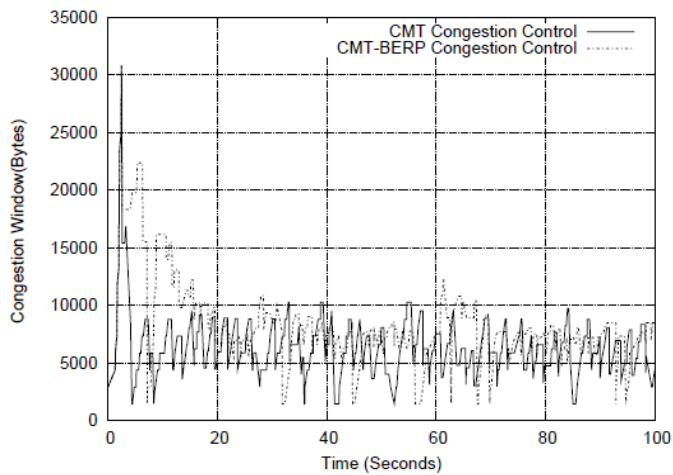

Figure 9. Congestion Window evolution for average UDP traffic load of 0.9.